\documentclass[11pt]{article}
\usepackage{amssymb}
\usepackage{graphicx,epsfig}

\begin{document}
\newcommand{\br}{\hbox{\boldmath$r$}}  
\newcommand{\ba}{\hbox{\boldmath$a$}}  
\newcommand{\bb}{\hbox{\boldmath$b$}}  
\newcommand{\bk}{\hbox{\boldmath$k$}}  
\newcommand{\bro}{\hbox{\boldmath$\rho$}}  
\newcommand{\ro}{\rho}  
\newcommand{\ru}{\ro(\varphi,w)}  
\newcommand{\ra}{\ro(\varphi_1,w_1)}  
\newcommand{\rb}{\ro(\varphi_2,w_2)}  
\newcommand{\rsf}{\ro_1^2(\varphi,w)}  
\newcommand{\rss}{\ro_2^2(\varphi,w)}  
\newcommand{\rta}{\ro(\theta,v)}  
\newcommand{\rtb}{\ro(\theta,w)}  
\newcommand{\om}{\Omega(x,y)}  
\newcommand{\oa}{\Omega_x}  
\newcommand{\ob}{\Omega_y}  
\newcommand{\ok}{\hat\Omega_{\bk}}  
\newcommand{\ou}{\hat\Omega_0}  
\newcommand{\deo}{\delta\Omega(x,y)}  
\newcommand{\dro}{\delta\ro(\varphi,w)}  
\newcommand{\tn}{\tan^{-1}\frac{y-\eta}{x-\xi}}  
\newcommand{\tno}{\tan^{-1}\frac{y'-\eta}{x'-\xi}}  
\newcommand{\ta}{\tan^{-1}\frac{\eta_1-\eta}{\xi_1-\xi}}  
\newcommand{\tb}{\tan^{-1}\frac{\eta_2-\eta}{\xi_2-\xi}}  
\newcommand{\di}{(x-\xi)^2+(y-\eta)^2}  
\newcommand{\dio}{(x'-\xi)^2+(y'-\eta)^2}  
\newcommand{\da}{(\xi_1-\xi)^2+(\eta_1-\eta)^2}  
\newcommand{\db}{(\xi_2-\xi)^2+(\eta_2-\eta)^2}  
\newcommand{\In}{\int_0^M\int_0^{2\pi}}  
\newcommand{\varfi}{\varphi}  
\newcommand{\si}{\sin\varphi d\varphi dw}  
\newcommand{\co}{\cos\varphi d\varphi dw}  
\newcommand{\xa}{\xi+\ro(\varphi,w)\cos\varphi}  
\newcommand{\ea}{\eta+\ro(\varphi,w)\sin\varphi}  
\newcommand{\xb}{\xi+\ro(\theta,v)\cos\theta}  
\newcommand{\eb}{\eta+\ro(\theta,v)\sin\theta}  
\newcommand{\xc}{\xi+r\cos\varphi}  
\newcommand{\ec}{\eta+r\sin\varphi}  
\newcommand{\dl}{(\lambda_1-\lambda_2)}
\newcommand{\Ru}{\ro_1^2+\ro_2^2-2\ro_1\ro_2\cos(\varphi_1-\varphi_2)}  
\newcommand{\Rc}{\ro_1^2+\ro_2^2-2\ro_1\ro_2\cos\varphi}  
\newcommand{\Rd}{r_1^2+r_2^2-2r_1r_2\cos(\varphi_1-\varphi_2)}  
\newcommand{\Aa}{\left.\frac{\delta
F}{\delta\Omega(x,y)}\right|_{x=\xi+\ro(\varphi,w)\cos\varphi,\ \
y=\eta+\ro(\varphi,w)\sin\varphi}}  
\newcommand{\Bb}{\left.\frac{\delta
F}{\delta\Omega(x,y)}\right|_{x=\xi+\ro(\theta,v)\cos\theta,\ \
y=\eta+\ro(\theta,v)\sin\theta}}  
\newcommand{\Cc}{\left.\frac{\delta
F}{\delta\Omega(x,y)}\right|_{x=\xi+r\cos\theta,\ \ y=\eta+r\sin\theta}}  

\begin{center}
{\Large \textbf{Hamiltonian description of vortex systems}}\\%
[1cm]
{\textbf{Leonid I. Piterbarg}}

{Department of Mathematics, University of Southern California}

{Kaprielian Hall, Room 108,  Los Angeles, CA 90089-2532, USA}

\end{center}

\bigskip

{\Large \textbf{Abstract }}

\bigskip

In the framework of 2D ideal Hydrodynamics a vortex system is  defined as a smooth vorticity function having few positive local maxima and negative local minima separated by  curves of zero vorticity. Invariants of such structures are discussed following from the vorticity conservation law and invertibility of Lagrangian motion. Hamiltonian formalism for vortex systems is developed by introducing new functional variables diagonalizing the original non-canonical Poisson bracket.

\bigskip


\bigskip

{\textbf{\Large {1. Introduction }}

\bigskip
Fundamentals of Hamiltonian formalism in application to hydrodynamics systems were developed in [1,2] and proved to be an efficient tool in handling a variety of fluid mechanics problems.
Here that approach is applied to a relatively simple object: a vortex system in the framework of ideal $2D$ hydrodynamics. Our starting point is the vorticity conservation equation  
$$
\frac{\partial\Omega}{\partial t}+J(\psi, \Omega)=0 
\eqno(1)  
$$  
 written  in the Hamiltonian form  as, [3]
$$ 
\frac{\partial\Omega}{\partial t}=\{H,\Omega \}\  ,
$$  
where the non-canonical Poisson bracket is expressible as
$$  
\{F,G\}_{\Omega}=\int\Omega(\br)J_{x,y}\left(\frac{\delta F}{\delta\Omega(\br)}, \frac{\delta
G}{\delta\Omega(\br)}\right)d\br\ . \eqno (2)  
$$  
Here $F=F(\Omega), G=G(\Omega)$ are smooth functionals, $\br=(x,y)$, and $J_{x,y}(f,g)=f_xg_y-f_yg_x$  the Jacobian.  Henceforth the integration region is the whole plane $R^2$  unless other indicated. The Hamiltonian is given by 
$$  
H=\frac 12\int|\nabla\psi|^2d\br=-\frac 12\int\psi\Omega d\br   \eqno (3)
$$
that implies
$$
\displaystyle \frac{\delta H}{\delta \Omega}=-\psi .\eqno (4)
$$
Combining (2) and (3) we extend (1) to any functional
$$
\frac{\partial F(\Omega)}{\partial t}=\int\frac{\delta F}{\delta\Omega(\br)}J_{x,y}\left(\psi,\Omega(\br)\right)d\br  . 
\eqno(5)  
$$  
Next we assume that $\Omega(\br)$ decays fast enough as $|\br|\to\infty$ and, hence, the stream function is expressed in terms of vorticity as
$$
\psi(\br) =\displaystyle \frac{1}{2\pi}\int\ln(\br-\br')\Omega(\br')d\br'. \eqno(6) 
$$
From (4) and (6) 
$$
H=\displaystyle \frac{1}{4\pi}\int\int \ln(\br_1-\br_2)\Omega(\br_1)\Omega(\br_2)d\br_1d\br_2 .  \eqno(7) 
$$
 The main objectives of this work are:

- Formulate and prove rigorously conservation laws for the topography of vorticity $\Omega$ such as the number of critical points (points where $\nabla\Omega =\bf 0$), the vorticity values at the critical points, and the number of distinct level curves defined by $\Omega (\br)=w$, corresponding to a fixed value $w$, that also will be called contour lines.

- Derive translation equations for the critical points

- Derive and discuss equations for contour lines in both a Hamiltonian form and in form of closed integro-differential equations. In this regard our efforts can be viewed as an extension of the contour dynamics, [4, 5], to smooth vorticity functions

- Demonstrate how some well-known models such as point vortex systems and FAVOR, [6], could be derived from the underlying vorticity class.

Now we specify the class of functions $\Omega (\br)$  called the $N$-vortex system. Let
$$
\mathbf H_{\Omega}(\br)=\left( 
\begin{array}{cccccc}
\Omega_{xx} & \Omega_{xy}  \\ 
\Omega_{xy} & \Omega_{yy} 
\end{array}
\right)
$$
be the Hessian of vorticity. Assume that

(i) $\Omega$ has exactly $N$ extrema (maxima or minima) at points ${\bf{z}}_k=(\xi_k,\eta_k),\ k=1,...,N$, i.e.
$$
\displaystyle \nabla\Omega({\mathbf z}_k)={\bf 0},\ \ \det \left( \mathbf H_{\Omega}({\bf{z}}_k)\right) >0 .
$$

(ii) The set $\Gamma_0=\{\br\in R^2|\  \Omega(\br)=0\}$ is either empty or divides the plane in  $N\geq 2$ distinct regions $G_k,\  k=1,...,N$ such that each $G_k$ contains exactly one extremum  and  the vorticity has the same sign for all points in $G_k$.
 Thus, summarizing
$$
R^2=\bigcup_kG_k,\ \ G_k\cap G_j=\varnothing,\ \ \ \Omega(\partial G_k)=0,\ \ \mathbf{z}_k\in G_k .
$$

(iii) For any two adjacent regions $G_k$ and $G_j$ the signs in $G_k$ and $G_j$ are opposite. 

The last condition can be ensured by assumption that the set $\Gamma_0$ of zero vorticity lines is an Euler graph [7].  For such graphs the corresponding regions can be painted by two colors only in a way that any two adjacent regions have opposite colors. Our 'colors' mean positive and negative vorticity. Examples of some important vortex systems satisfying to  (i-iii) are shown in Fig.1. 
\begin{figure}[tbph]
\begin{center}
\includegraphics[width=1.6in, height=1.6in]{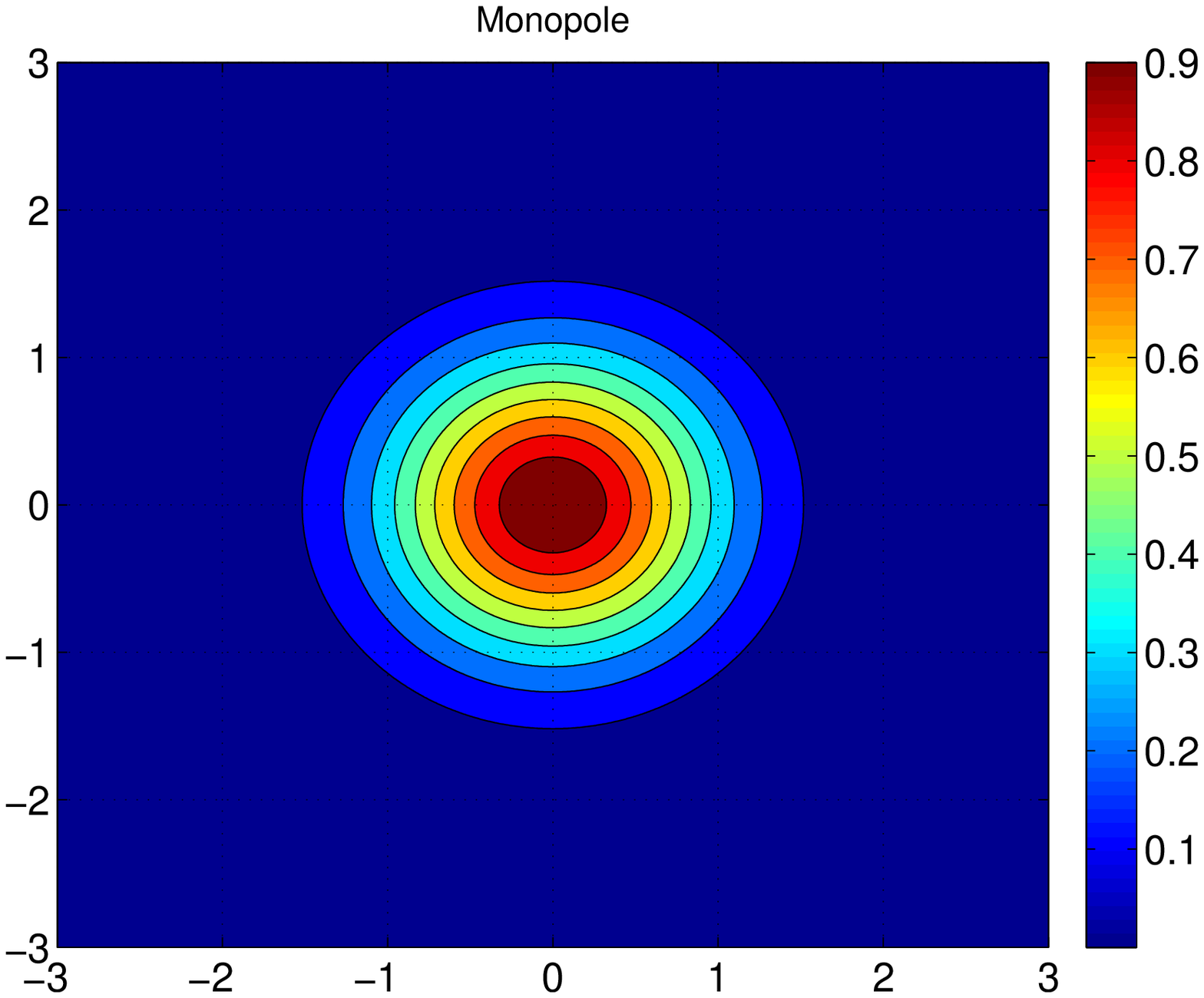} %
\includegraphics[width=1.6in, height=1.65in]{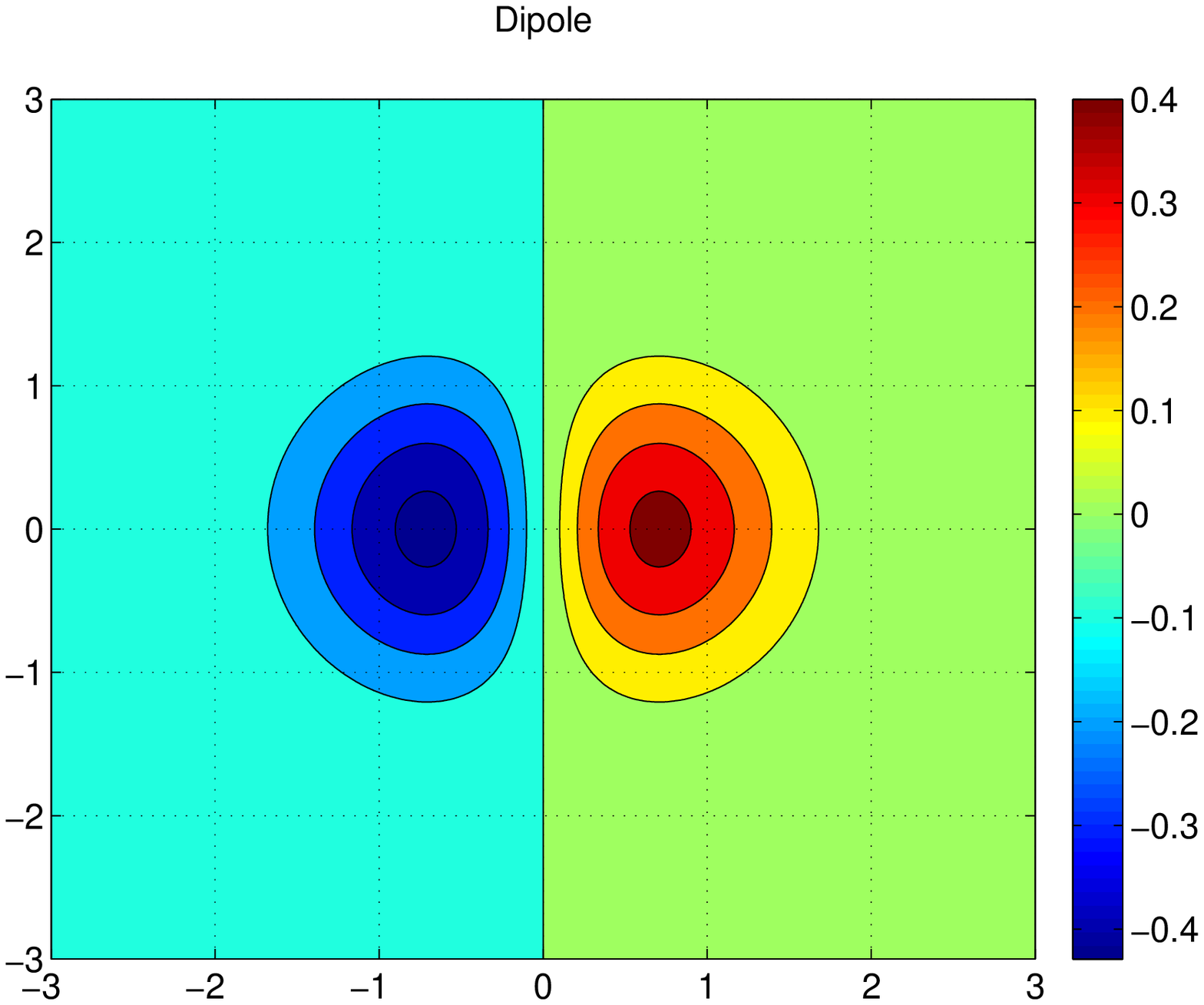}
\includegraphics[width=1.6in, height=1.6in]{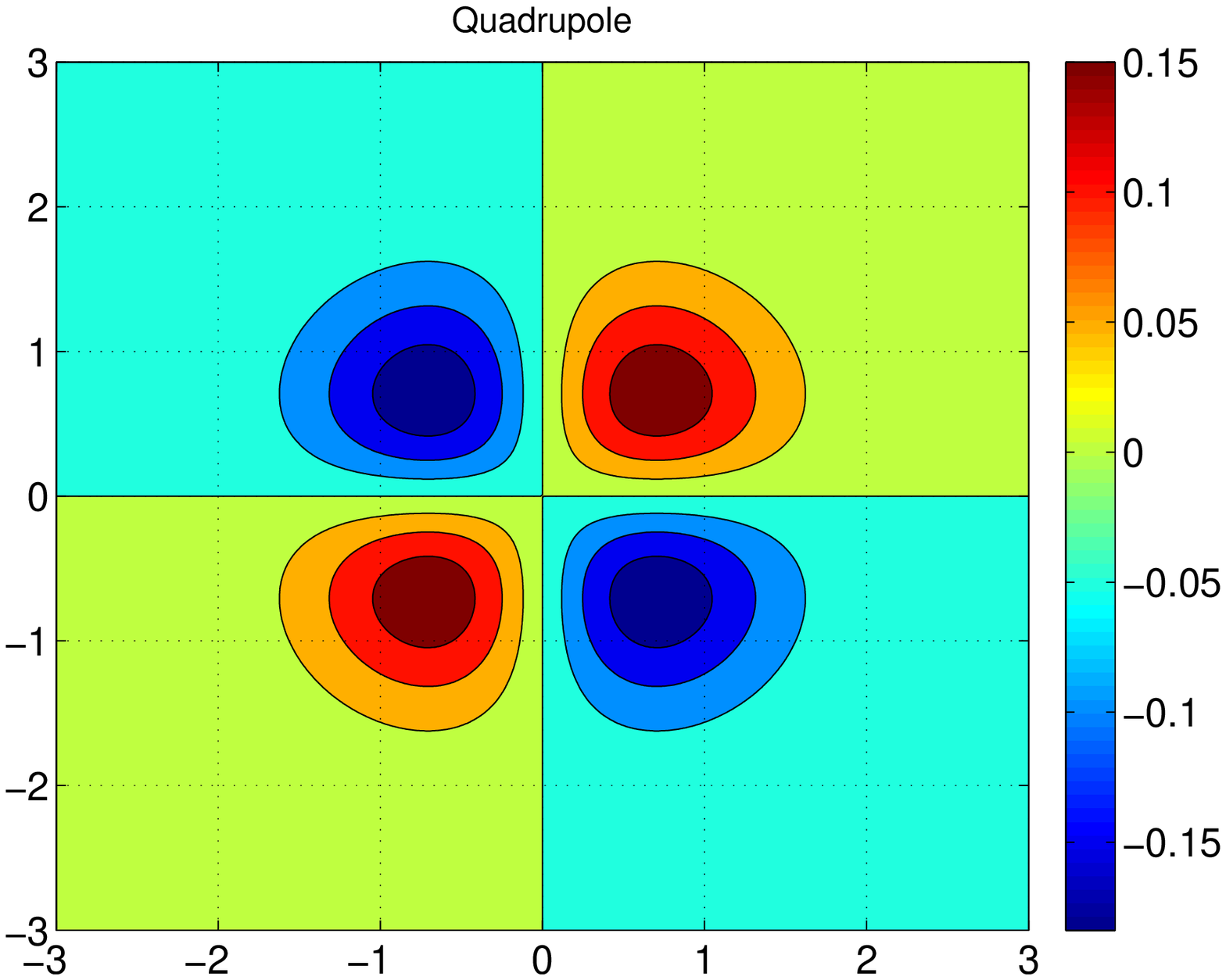}
\includegraphics[width=1.6in, height=1.6in]{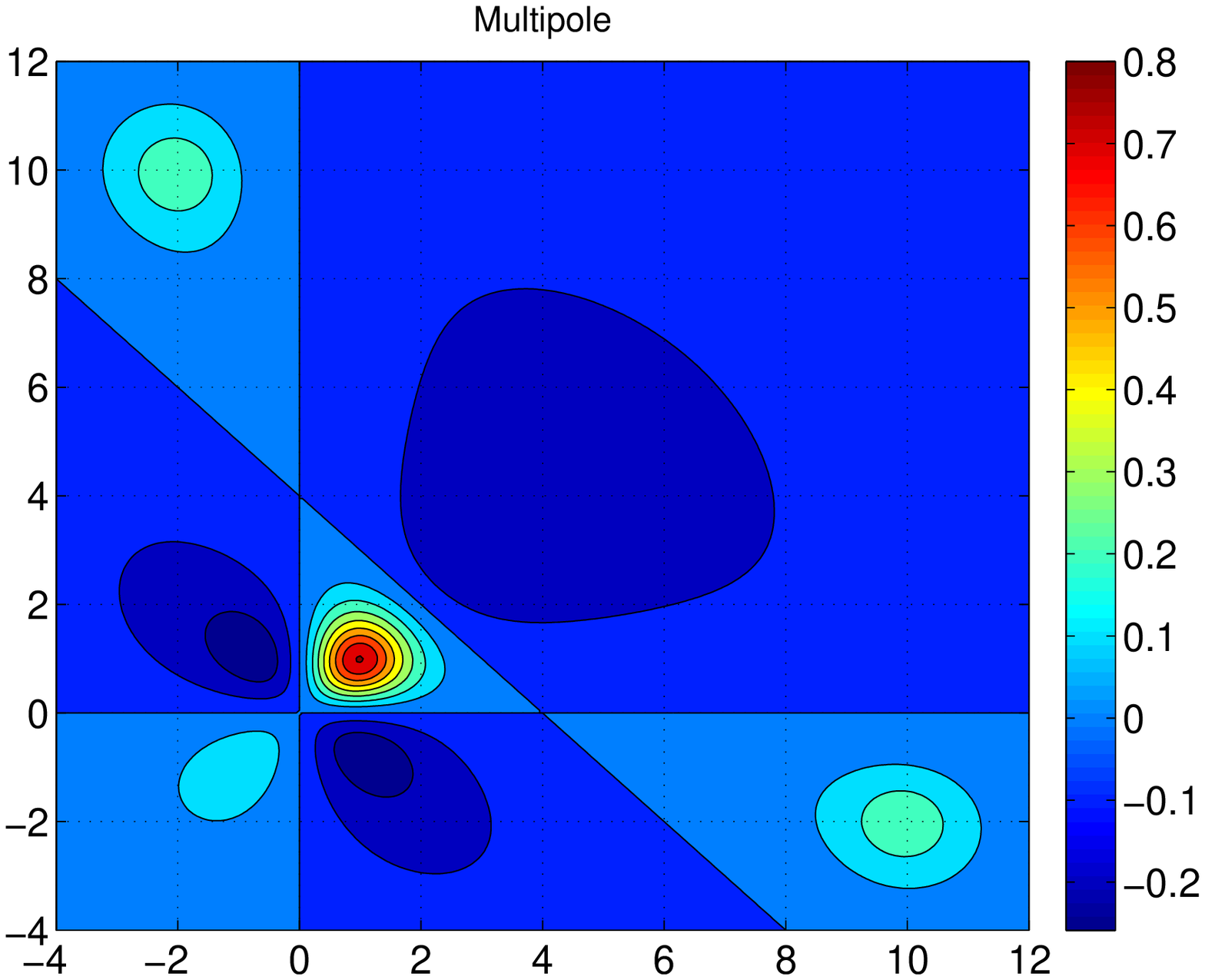}\\
\par
\smallskip
\end{center}
\par
\noindent\textsf{Figure 1.Examples of $N$ vortex systems }
\end{figure}

A remarkable fact proven in the text is that the structure (i-iii) and, in particular, the value of $N$ is conserved by equation (1). In other words critical points neither are being  born nor dying during the system evolution. Moreover, the vorticity values at the critical points
$
\omega_k= \Omega(\mathbf{\zeta}_k)
$
are conserved as well.

We do not address specifically saddles of $\Omega(\br)$ (hyperbolic critical points)  as not important in context of our goals.  However, it is worth noting  that their number and vorticity values are also conserved. We can say even more  assuming that the Euler graph representing $\Gamma_0$ has exactly four edges incident to each vertex like in two last panels of Fig.1. Namely, in this case each vertex (an intersection of  two zero vorticity lines) is a saddle  and vice verse: each saddle is a vertex of the graph. Thereby the vorticity at each saddle is zero.   
There is one more important invariant that is a  consequence of two fundamental laws: vorticity conservation in Lagrangian particles and incompressibility. For a fixed $w>0$ let $n(w)$ be the number of disjoint connected regions where $\Omega>w$ and the number of disjoint connected regions where $\Omega<w$ if $w<0$. We will show that  $n(w)$ is also conserved by (1). In other words, no merging of vorticity patches is possible in ideal $2D$ hydrodynamics. We take an opportunity to clearly spell out that obvious claim because in last years the vortex merger problem was addressed in many works, i.e. [8], but not always  initial equations were formulated explicitly. 

An important role of the extreme points is that they serve as natural poles for local polar coordinates in parametrization of vorticity lines at each particular region $G_k$. Let $r=\rho_k(\varphi,w)$ be the polar equation of the single contour line corresponding to the vorticity level $w$ at a certain region 
$$
\Gamma_{w,k}=\{\br\in G_k|\ \ \Omega(\br)=w)\} .\eqno (8)
$$

We derive closed evolution equations for $\rho_k,\ \ k=1,...,N$ that explicitly show interacting between vorticity contours corresponding to different regions $G_k$ and different vorticity levels $w$.  An essential drawback of polar parametrization is that the very assumption that curve can be represented in polar form with a single valued $\rho(\varphi)$ is not conserved by the system. Thus such a consideration is valid for small integration times only or if one considers a small vicinity of an extremum point or, equivalently, small values of $\rho_k$  where curve (8) is well approximated by an ellipse. 
This is why we first address a general parametrization (still pinned to $G_k$),  $\ \br=\hat{\br}(p,w)$ of $\Gamma_{w,k}$ where $p$ is a positive parameter, say the length of arc (natural parametrization). Such a parametrization  is free from the above drawback.

In both cases, polar and natural parametrization, the resulting equations are too sophisticated (non-linear, integro-differential, non-decoupling) to efficiently work on them. Exceptions are  monopoles and dipoles ( $N=1$, $2$ respectively). Thus, for now the equations describing evolution of contours (8) for $N>2$ are of pure theoretical interest only: our point is to clarify conservation laws concerning with vorticity topography and to shed light on the underlying Hamiltonian structure rewritten in new phase variables $\{\rho_k(\varphi,w)\}$ or $\{\hat{\br}_k(p,w)\},\ \ k=1,..., N$

Worth noting that condition (iii) is not necessary for deriving the mentioned equations. Its intention is to exclude unstable vortex structures once and forever. For the same reason we do not address bifurcation points where the determinant of Hessian is zero.

The paper is organized as follows. In  Section 1  we state and give a sketch of proof of most important conservation laws for  the considered vortex systems.  In Section 2 the main result on the Poisson bracket diagonalization is proven. As a consequence, equations for vorticity lines and extrema translation are derived for a general parametrization. These equations are specified and discussed for the polar parametrization in Section 3. Examples of application of the equations to monopoles and dipoles are given in Section 4. Section 5 contains a short discussion and conclusions. Finally, some details are brought to Appendix.

\bigskip

{\textbf{\large {1. Invariants }}

\bigskip

{ {PROPOSITION 1.\ } {\it{The number of critical points and their type is conserved by equation}} (1).

\smallskip

{{PROOF}}. Let $\br (t,{\bf {r_0}})=\br$ be the position of a Lagrangian particle at moment $t$ starting from $\bf r_0 $ and  $\bf J(\br, \bf r_0)=\partial (\br)/\partial (\bf r_0)$ be the Jacobi matrix of the diffeomorphism $T: \br_0\rightarrow \br$.  Direct computations based on the vorticity conservation equation (1) give 
$
\nabla \Omega (\br)={\bf {J}} \nabla\Omega({\bf{ r}}_0)$ and ${\mathbf {H}}_\Omega (\br)= {\bf {J}}{\mathbf {H}}_\Omega (\br_0){\bf {J}}^*$ if   $\nabla\Omega(\bf r_0)=\bf 0$, the star means transposition. Now the statement  follows from $\det\left(\mathbf J\right)=1$ that is a consequence of incompressibility, [9].

Due to  assumptions (i-ii) for any $w\neq 0$ the set $\Gamma_w=\{\br | \Omega(\br)=w\}$ consists of a finite number of closed curves  and  the number $n(w)$ of such curves  obviously does not exceed $N$. The next statement shows that $n(w)$ is conserved as well.

\bigskip

{{PROPOSITION 2}} {\it{Assume for definiteness that}} $w>0$ {\it{and at the initial moment region}} $D=\{\br  | \Omega(\br)>w\}$ = $D_1\cup D_2$ {\it { consists of two  disjoint regions. Then}} $D$ {\it{remains so for all moments.}}

\smallskip

{{PROOF}}. Assume that at some moment $t$ the regions have merged.  Let $\br_1\in D_1$ and $\br_2\in D_2$. Consider  a continuous curve $C$ joining $T(\br_1)$ and $T(\br_2)$ completely belonging to the merger, i.e. $C\in T(D_1)\cup T(D_2)$.  That means the vorticity at each point of $C$ is greater than $w$,
$\Omega (C)>w$. Thus $\Omega (T^{-1}(C))>w$ as well, but the curve $T^{-1}(C)$  for sure lies partially outside both $D_1$ and $D_2$ and hence at some point of the curve $\Omega <w$ that is a contradiction. By invertibility of $T$ a connected region cannot be broken down in two distinct ones either.

\bigskip

{{PROPOSITION 3}} {\it{The local extrema are conserved, more exactly the equations}}
$$
\frac{\partial \xi_k}{\partial t}=-\psi_y(\xi_k, \eta_k),\ \ \frac{\partial \eta_k}{\partial t}=\psi_x(\xi_k, \eta_k) \eqno (9)
$$
{\it{hold true for any}} $k=1,..,N$

\smallskip

{{PROOF.}} Let us fix a certian $G_k$ and  drop  sub $k$ in the following below computations for simplicity. With the assumption that there is a single critical point of $\Omega$ in $G$ we can represent its coordinates as the following functionals of $\Omega$ , [10]
$$
\xi(\Omega)=\int_{G}x\delta(\Omega_x)\delta(\Omega_y)S(\Omega)d\br,\ \ \eta(\Omega)=\int_{G}y\delta(\Omega_x)\delta(\Omega_y)S(\Omega)d\br\ , \ \ \ 
$$
where $S(\Omega)=\Omega_{xx}\Omega_{yy}-\Omega_{xy}^2$ is the determinant of the Hessian. The idea behind such a representation is that a unique solution of 
$$
\mathbf f (\mathbf r)=\mathbf 0,\ \ \ \mathbf r \in R^n,\ \ \mathbf f: R^n\rightarrow R^n \eqno(10)
$$
is expressible as
$$
\mathbf r_0 =\displaystyle \int_{R^n}\br \delta(\mathbf f (\mathbf r))\left|\det\left(\frac{\partial \mathbf f}{\partial \mathbf r}\right)\right|d\mathbf r , 
$$
where $\partial \mathbf f/\partial \mathbf r$ is the Jacobi matrix of map (10).

Taking variational derivatives in (9) obtain, [10]
$$
\displaystyle \frac{\delta \xi}{\delta\Omega(x,y)}=\delta'(\Omega_x)\delta(\Omega_y)S(\Omega),\ \ 
\displaystyle \frac{\delta \eta}{\delta\Omega(x,y)}=\delta'(\Omega_y)\delta(\Omega_x)S(\Omega),\ \  
$$
and plug each expression in (5). After changing the variables in the integrals $u=\Omega_x(x,y)$ and $v=\Omega_y(x,y)$ 
arrive at (9).
\bigskip

{\textbf{\large {2. Poisson bracket diagonalization. Evolution and translation equations for vorticity lines}}

\bigskip

 Assume that vorticity values at the extrema are ordered $\omega_1>\omega_2>...>\omega_N$. 
Let 
$$
 x=\xi_k+\hat{x}_k(p,w),\  y=\eta_k+\hat{y}_k(p,w),\ \ \ (p,w)\in D_k = \{ 0\leq p < L_k(w),\ 0<w<\omega_k\} \eqno (11)
$$ 
be an arbitrary parametrization of the vorticity line $\Gamma_{w,k}$ (see (8)), corresponding to level $w$  in region $G_k$, $p$ is a positive  parameter with the upper limit $L_k(w)$ depending in general on $w$. For example, in the case of natural parametrization it is the length of a particular contour. According to (11), the origin of local rectangular coordinate system $(\hat x,  \hat y)$ is placed at ${\bf{z}}_k=(\xi_k,\eta_k)$. 

\bigskip

{ {ASSUMPTION}}  {\it{The map}} $(p,w)\rightarrow (x,y)$ {\it{defined by }}(11) {\it{is one-to-one map of $D_k$ onto $G_k$ for a fixed $(\xi_k,\eta_k)$}} .

\smallskip

The assumption implies that the inversion  of (11) can be written as
$$
p=p_k(x,y),\ \ w=\Omega(x,y) ,
$$
where  $p_k(x,y)$  is a function determined by a specific parametrization and it depends on region $G_k$.
Introduce  ${\bf e} = (1,1)$,
$$
\mathbf V(\br)=\frac{1}{S(\Omega)}\left( 
\begin{array}{cccccc}
-\Omega_{xy} & \Omega_{xx}  \\ 
\Omega_{yy} & -\Omega_{yx} 
\end{array}
\right),\ \ \hat{\br}_k (p,w)=(\hat{x}_k(p,w), \hat{y}_k(p,w)),\ \ \nabla \delta (\br)=(\delta'(x)\delta(y), \delta(x)\delta'(y))^* .
$$

The following statement proven in Appendix plays a key role in further computations. 

\bigskip

{{LEMMA. \ }} {\it{Consider the vector function}} $\hat{\bf r}(p,w)$ {\it{ of local coordinates on $\Gamma_{k,w}$ in $G=G_k$  as a functional of}} $\Omega(\br )$,  {\it {  then its variational derivative is given by}} 
$$
\begin{array}{cc}
\displaystyle \frac{\delta \hat{\bf r}(p,w)}{\delta\Omega(\br)}= \frac{\hat{\bf r}_w(p,w)}{g(p,w)}\left(\delta(\Omega(\br)-w)\delta(p(\br)-p)-{\bf e}\frac{\partial}{\partial p}\left(\hat{\bf r}(p,w)\mathbf V(\bf z)\nabla \delta(\br -\bf z)\right)\right) .
\end{array}\eqno (12)
$$
{\it {where}} $g(p,w)=\hat{x}_p\hat{y}_w-\hat{y}_p\hat{x}_w$

\smallskip
For brevity   we dropped subscripts $k$ in (12).
Set
$$
\zeta_k(p,w)=\int_w^{\omega_k} g_k(p,u)du  \eqno (13)
$$
and let $F=F(\zeta_1,...,\zeta_N)$ be a smooth functional of new variables. Obviously, it is also a functional of $\Omega$ that will be denoted by the same symbol $F(\Omega)$.  Next, introduce the range $R_k$ of values of $\Omega (\br),\ \ \br\in G_k$ that is either interval $(0,\omega_k)$ or $(\omega_k,0)$ depending on the sign of $\omega_k$.  Finally,  let $S(w)=\{\left.k\in\{1,2,...,N\}\right|\ \ w\in R_k\}$ be the list of all regions containing a piece of $\Gamma_{w}=\{\br |\  \Omega (\br) =w\}$.

\bigskip

{{PROPOSITION 4}} 
{\it{ Assume that}} $\delta F/\delta \Omega(\br)$ {\it{is a smooth function of}} $\br$, {\it{then}}
$$
\displaystyle  \frac{\delta F}{\delta \zeta_k(p,w)} =\left.\frac{\delta F}{\delta \Omega(\br)}\right|_{\br={\bf z}_k+\hat{\br}_k(p,w)}
$$
{\it{for}} $w \neq \omega_k$ {\it{and}}

$$
 \{\zeta_k(p,w),F\}=L_k(F) ,\eqno (14)  
$$
{\it{where}}
$$
\displaystyle  L_k(F)=\frac{\partial}{\partial p}\left\{ \sum_{j\in S(w)}\left(\left.\frac{\delta F}{\delta\zeta_j(q,w)}\right|_{q=p_j(\Delta {\bf z}_{kj}+\hat{\bf{ r}}_k(p,w))}\right)-
 \nabla\left.\frac{\delta F}{\delta\Omega(\br)}\right|_{\br ={\bf z}_k}\cdot\hat{\bf{ r}}_k(p,w)\right\},\ \ \Delta {\bf z}_{kj}={\bf z}_{k}-{\bf z}_{j} .
$$
 
\smallskip

Notice that the Poisson bracket $\{\zeta_k(p,w), \zeta_j(q,u)\}$ is not defined for $k=j$.

Main steps in derivation of (14) are as follows. First, using (13) , the chain rule, and
$$
\displaystyle \frac{\delta \zeta(p,w)}{\hat{\br}(q,u)}=\delta(p-q)\delta(w-u)\hat {\br}^\perp_p(q,u)-I_{(w, \omega_k)}(u)\delta'(p-q)\hat {\br}^\perp_u(q,u)\ ,
$$
 where $I_A(x)$ is the indicator of $A$  and $(x,y)^\perp=(-y,x)$, we get
$$
\begin{array}{cc}
\displaystyle \frac{\delta \zeta(p,w)}{\delta\Omega(\br)}= \delta(\Omega(\br)-w)\delta(p(\br)-p)-\frac{\partial}{\partial p}\left(\hat{\bf r}(p,w){\bf V}\nabla \delta(\br -\bf z)\right) ,
\end{array}
$$
that leads to the first statement.
Then plug the obtained expression in (2) and break down the integration over $R^2$ in integration over $G_j,\ j=1,...,n(w)$. Finally we proceed to local coordinates $(\hat x_j, \hat y_j) $.

Thus, setting in (14) $F=H$ we get  Hamiltonian equations for the new  variables 
$$
 \frac{\partial \zeta_k(p,w)}{\partial t}=L_k(H) .
\eqno(15)
$$
To get a closed system in terms of variables $\zeta_k$  we  again change the integration over the whole plane in (7) with integration over distinct $G_k,\ k=1,2,..., N$. The result is $H=\displaystyle \sum_{k,j=1}^N H_{kj}$ where
$$
\begin{array} {cc}
 \displaystyle H_{kj}=\frac{s_ks_j}{4\pi}\int_0^{\omega_k}\int_0^{\omega_j} \int_0^{L_k(w_1)} \int_0^{L_j(w_2)}w_1w_2\zeta_k(p_1,w_1)_{w_1}\zeta_j(p_2,w_2)_{w_2} \ln D_{kj}(p_1,w_1,p_2,w_2)dp_1dp_2dw_1dw_2 ,
\end{array}  \eqno(16) 
$$
where
$$
s_k=sgn(\omega_k),\ \ D_{kj}=\sqrt{(\Delta\xi_{kj}+\hat{x}_k(p_1,w_1)-\hat{x}_j(p_2,w_2))^2+(\Delta\eta_{kj}+\hat{y}_k(p_1,w_1)-\hat{y}_j(p_2,w_2))^2} ,
$$
$\Delta\xi_{kj}=\xi_k-\xi_j,\ \ \Delta\eta_{kj}=\eta_k-\eta_j$.  A cumbersome expression for Hamiltonian is a trade-off for a diagonal Poisson bracket. 
Finally,  $\hat x, \hat y$ should be  expressed in terms of $\zeta$.  That can be easily done in the case of a polar parametrization we consider below.

\bigskip

{\large{\bf{3. Polar parametrization}}}

\smallskip

Assume now $p=\varphi$ is a polar angle and for each $G_k$ introduce local polar coordinates
$$
 x=\xi_k+\rho_k(\varphi,w)\cos \varphi,\  y=\eta_k+\rho_k(\varphi,w)\sin \varphi,\ \ \ (\varphi,w)\in D_k=[0,2\pi]\times [0,\omega_k] ,\ \ (x,y)\in G_{k} ,
$$ 
where $\rho_k(\varphi,w)$ is the distance from ${\bf{z}}_k$ to the point on the contour in direction $\varphi$.  In other words, the closed curve $\Gamma_{w,k}$ is covered by equation $r=\rho_k(\varphi,w)$. Easy to see that the new phase variable introduced in (13) now becomes $\zeta_k =\rho_k^2(\varphi, w)/2$ and (15) implies

\bigskip

{{POPOSITION 5}} 
$$
\begin{array}{lll}
\displaystyle \frac 14\frac{\partial \rho^2_k(\varphi,w)}{\partial t}=\frac{\partial}{\partial\varphi}\left\{\frac{\delta H}{\delta\rho^2_k(w,\varphi)}+\sum_{j\neq k}\frac{\rho_j(\theta_{kj},w)}{\rho_k(\varphi,w)}\left.\frac{\delta H}{\delta\rho^2_j(\theta,w)}\right|_{\theta=\theta_{kj}(\varphi)}-\rho_k(\varphi,w)\left(D_\varphi\frac{\delta H}{\delta\rho^2_k}\right)_{w=\omega_k}\right\} ,
\end{array}\eqno (17)
$$
{\it{where}} $D_\varphi$ {\it{is the derivative in direction given by}} $\varphi$ {\it{and}}\  $\theta =\theta_{kj}(\varphi)$ {\it{is the solution of}}
$$
\displaystyle \frac{\Delta\eta_{kj}+\rho_j(\theta,w)\sin\theta}{\Delta\xi_{kj}+\rho_j(\theta,w)\cos\theta}=\tan\varphi .
$$
\smallskip
The last term in the braces reflects an effect on the shape of $k$-th vorticity line due to the vortex peak motion.

Notice that for the polar coordinates (16) turns into
$$
\begin{array}{cc}
 H_{kj}=\displaystyle \frac{s_ks_j}{16\pi}\int_0^{\omega_k}\int_0^{\omega_k}\int_0^{2\pi} \int_0^{2\pi}w_1w_2\left(\rho_k^2 \right)_{w_1}\left(\rho_j^2 \right)_{w_2}\ln D_{kj}(\varphi_1,w_1,\varphi_2,w_2)d\varphi_1d\varphi_2dw_1dw_2 ,
\end{array}
$$ 
{\it{where}} 
$$\rho_k=\rho_k(\varphi_1,w_1),\ \rho_j=\rho_j(\varphi_2,w_2),\ \ \ D_{kj}=\sqrt{(\Delta\xi_{kj}+\rho_k\cos\varphi_1-\rho_j\cos\varphi_2)^2+(\Delta\eta_{kj}+\rho_k\sin\varphi_1-\rho_j\sin\varphi_2)^2}
$$

To get a closed system for $\rho_k$ one should rewrite (17) in terms of stream function $$\psi_j(\theta,w)= \psi\left(\xi_j+\rho_j(\theta,w)\cos \theta, \eta_j+\rho_j(\theta.w)\sin \theta\right).$$ 
The result is
$$
\begin{array}{lll}
\displaystyle \frac 12\frac{\partial \rho_k(\varphi,w)}{\partial t}=-\frac{1}{\rho_k(\varphi,w)}\frac{\partial}{\partial\varphi}\left\{\psi_k \left(\varphi,w\right)+ \sum_{j\neq k} \psi_j\left(\theta_{kj},w\right)-\rho_k(\varphi, w)D_\varphi\psi(\varphi,\omega_k)\right\} ,
\end{array}
$$
and then substitute for $\psi$  the following expression
$$
\begin{array}{ccc}
\displaystyle \psi_j(\theta,w)=-\frac{1}{2\pi}\sum_{\alpha=1}^{N}
s_\alpha\int_{0}^{2\pi}\int_0^{\omega_\alpha}u\rho_\alpha(u,\theta)(\partial \rho_\alpha(u,\theta))/\partial u)\ln D_{j\alpha}(\theta, w, \varphi,u)dud\varphi 
\end{array} , \eqno (18)
$$
derived from (8).

Translation equations for the critical points we give in terms of complex coordinates $z_k=\xi_k+i\eta_k$ by plugging (18) in (9)
$$  
\begin{array}{l}  
\displaystyle\frac{\partial z_k^*}{\partial
t}=\displaystyle\frac{s_k}{2\pi i}\int_0^{\omega_k}\int_0^{2\pi}\rho_k(\varphi, w)e^{-i\varphi} d\varphi
dw+ \sum_{j\neq k}\frac{s_j}{2\pi i}\displaystyle\int_0^{\omega_j}\int_0^{2\pi}
\frac{w\rho_j(\varphi, w)\left(\partial\rho_j(\varphi, w)/\partial w) \right)dwd\varphi }{z_k-z_j-\rho_j(\varphi, w)e^{i\varphi}}\ .
\end{array}
$$ 

Finally, let us express other well known invariants of (1) in terms of the new variables.  First, notice  Casimir functionals  $K(\Omega)=\int K\left(\Omega(\br)\right)d\br$, where $K(\cdot)$ is an arbitrary smooth function.  Breaking the integral over $R^2$ in regions $G_j,\ j=1,...,N$ and proceeding to new variables $(\xi_j, \eta_j, \rho_j)$  one gets
$$
\displaystyle K=-\sum_j\frac {s_j}{2}\int_0^{\omega_j}K(w)\frac{\partial}{\partial w}\left(\int_0^{2\pi}\rho^2_j(\varphi,w)d\varphi\right) dw .
$$
Notice that the inner integral is simply the doubled area of the region bounded by $\Gamma_{w,j}$.

Then, in the same manner the vorticity first momentum
$$
c=\int(x+iy)\Omega(\br)d\br
$$ 
can be obtained
$$
\displaystyle c=\sum_js_j\int_0^{\omega_j}\int_0^{2\pi}\left( \frac 12 z_j\rho^2_j(\varphi,w)+\frac 13 \rho^3_j(\varphi,w)e^{i\varphi}\right)d\varphi dw .
$$
\bigskip

{\textbf{\large {4. Monopole and Dipole}}

\bigskip
A monopole is defined by conditions  $N=1,\ \ \Omega (\br)>0,\ \ \br\in R^2$. Denote $M=\omega_1>0$ the maximum value of the vorticity and introduce local polar coordinates with pole at the point of maximum ${\bf {z}}=(\xi,\eta)$.

 In other words, the closed curve $\Gamma_w$ is covered by equation $r=\rho(\varphi,w)$ in local polar coordinates. As it was already noticed  $\rho(\varphi,w)$ does not remain a single valued function in the process of evolution except trivial cases such as circular contours for all $w$.
Thus, if we interpret  $\rho(\varphi,w)$ as a distance, then all the following equations hold true only during finite time (probably small) of integration. However, if we treat $\rho(\varphi,w)$  as a generalized distance ( a pseudo inverse of $\Omega$ with respect to radial variable $r$) i.e.
$$
\displaystyle \rho(\varphi,w)=\int_0^\infty I_{(0,\infty)}\left(\Omega(\xi+r\cos\varphi, \eta+r\sin\varphi)-w\right)dr\ , \eqno(19)
$$
then  the following below evolution equations hold true for all $t$. This is  because their derivation is based on the variational derivative of $\rho$ in $\Omega$ obtained from (19) rather than on the distance interpretation where it is assumed that the ray from ${\bf{z}}$ in direction $\varphi$ intersects $\Gamma_w$ once.  If it intersects the contour  few times, then $\rho$ given by (19) is the sum of the distances to all the intersection points.

In the considered case (17) becomes

$$
\frac 14\frac{\partial\rho^2(\varphi, w)}{\partial t}=\frac{\partial}{\partial \varphi}\left\{\frac{\delta H}{\delta\rho^2(\varphi,w)} -\rho(\varphi,w)\left(D_\varphi\frac{\delta H}{\delta\rho^2(\varphi,w)}\right)_{w=M}\right\} , \eqno (20)
$$
{{where}} $\rho=\rho(\varphi,w)$ {{and}}
$$
H=\displaystyle \frac{1}{16\pi}\int_0^M\int_0^M \int_0^{2\pi} \int_0^{2\pi}w_1w_2\left(\rho_1^2 \right)_{w_1}\left(\rho_2^2 \right)_{w_2}\ln D(\varphi_1,w_1,\varphi_2,w_2)d\varphi_1d\varphi_2dw_1dw_2 
$$ 
{{with}} 
$$
\rho_1=\rho(\varphi_1,w_1),\ \rho_1=\rho(\varphi_2,w_2),\ \ \ D=\sqrt{(\rho_1\cos\varphi_1-\rho_2\cos\varphi_2)^2+(\rho_1\sin\varphi_1-\rho_2\sin\varphi_2)^2} .
$$
Equation (20) was first appeared  in [10] where the singular term, describing effects of the vortex motion on its shape,  was missing. Yet, here the expression for Hamiltonian is significantly simplified. 

The translation equation  in terms of the complex coordinate $z=\xi+i\eta$ becomes
$$  
\begin{array}{l}  
\displaystyle\frac{\partial z^*}{\partial t}=\displaystyle\frac{1}{2\pi i}\int_0^M\int_0^{2\pi}\rho(\varphi, w)e^{-i\varphi} d\varphi dw  .
\end{array} 
\eqno(21)  
$$

Notice a similarity with the contour dynamics, [4],  where the object of study was a vorticity patch of value $\Omega (\br)=M$ bounded by a closed curve $ r=\rho(\varphi) $  with  zero vorticity outside. Not difficult to show that in this case (1) again leads to a Hamiltonian system resulting in the following evolution equation
$$
\frac 12\displaystyle \frac{\partial}{\partial t}\rho^2(\varphi) = \frac{2}{M}\frac{\partial}{\partial \varphi}\frac{\delta H}{\delta\rho^2}=-\frac{\partial}{\partial \varphi}\psi\left(\varphi\right) ,
 \eqno (22) 
$$
where $\psi\left(\varphi\right)=\left.\psi(\varphi, r)\right|_{r=\rho(\varphi)}$ and $\psi(\varphi, r)=\psi\left(\xi+r\cos\varphi,\eta+r\sin\varphi)\right)$. In this case the pole $(\xi,\eta)$ is usually placed at the patch centroid. It is now easy to get a closed equation for $\rho(\varphi)$ from (22) by using
$$
\displaystyle \psi\left(\varphi,r\right)=-\frac{M}{4\pi}\int_0^{2\pi}\left[\rho^2(\theta)-r\left(\rho(\theta)\sin(\theta-\varphi)\right)_\theta\right]\ln\left(r^2+\rho^2(\theta)-2r\rho(\theta)\cos(\theta-\varphi)\right)d\theta . \eqno(23)
$$
 Strange enough, we could not find in the literature an absolutely correct closed equation obtained from (22) after differentiating in the right hand side. For example, in [4] and [12] both, the expression (23) and the equation (22), involving stream function, were correct, but a mistake was made when differentiating stream function in $\varphi$.

Returning to  the  case of smooth vorticity notice that the stream function expression in coordinates $(\varphi, w)$
$$
\displaystyle \psi(\varphi,w)=-\frac{1}{4\pi}\int_0^{2\pi}\int_0^{M}u\rho(\theta,u)\rho_u(\theta,u)\ln\left(\rho^2(\varphi,w)+\rho^2(\theta,u)-2\rho(\theta,u)\rho(\varphi,w)\cos(\theta-\varphi)\right)dud\theta 
$$
is somewhat simpler than (23) because $\varphi$ shows up only under the logarithm. Another advantage of a continuous monopole compared to a patch is that the maximum (the vortex head) moves along stream lines while the centroid of patch certainly does not. 

Summing up,  a closed equation for $\rho(\varphi,w)=\rho(t,\varphi,w)$ can be written in form
$$
\displaystyle \frac{\partial}{\partial t}\rho^2=\frac{\partial}{\partial\varphi}N(\rho^2),\ \ \rho^2|_{t=0}=p(\varphi,w) , \eqno(24)
$$
where $p(\cdot, \cdot)$ is an initial condition and a non-linear integro-differential operator is
$$
\displaystyle N(\rho^2)(\varphi, w)= \frac{1}{2\pi}\int_0^{2\pi}\int_0^{M}\left(u\rho^2(\theta,u)_u \ln D  + 4\rho(\theta, u)\rho(\varphi, w)\cos(\varphi-\theta)\right)dud\theta
$$
with
$
D=\rho^2(\varphi,w)+\rho^2(\theta,u)-2\rho(\theta,u)\rho(\varphi,w)\cos(\theta-\varphi).\ 
$

Obviously, in general the equation (24) is not simpler than the original equation (1). However, in one particular case, we are about to discuss, making  use of (24) is more efficient than that of (1). Namely, we suggest a natural asymptotic procedure bridging the contour dynamics and the smooth vorticity case. For that let us assume that the initial vorticity for set up (1)  is represented as
$
\Omega(r,\varphi)=\displaystyle MS\left(\frac{r}{R(\varphi)}\right) 
$
where dimensionless function $S(x)$ defined on $[0,\infty)$ satisfies :  $S(0)=1,\ \ S'(x)<0,\ \ S(\infty)=0$ and $R(\varphi)$ is a certain space scale depending on the direction.
Introduce the following scaling
$$
\Omega_{\epsilon}(r,\varphi)=\displaystyle MS\left(\left[\frac{r}{R(\varphi)}\right]^{1/\epsilon}\right) , \eqno (25)
$$
that for small $\epsilon$ converts a continuously distributed vorticity to a patch
$$
\displaystyle \lim_{\epsilon\to 0}\Omega_{\epsilon}(r,\varphi)=\left\{
\begin{array}{ccc}
M,\ \ \ r<R(\varphi)\\
0, \ \ \ r>R(\varphi)
\end{array}
\right. 
$$
\begin{figure}[tbph]
\begin{center}
\includegraphics[width=2in, height=2in]{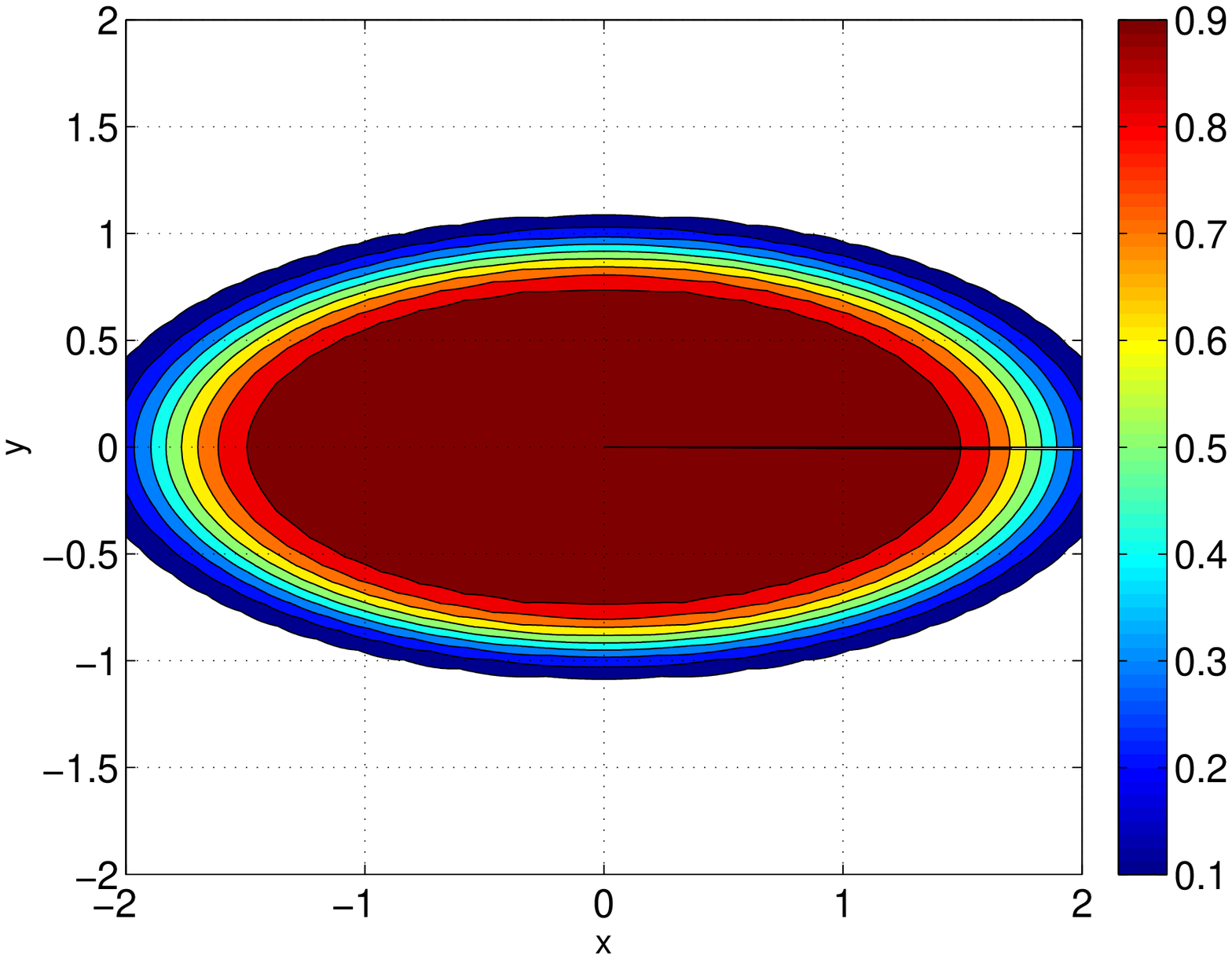}
\includegraphics[width=2in, height=2in]{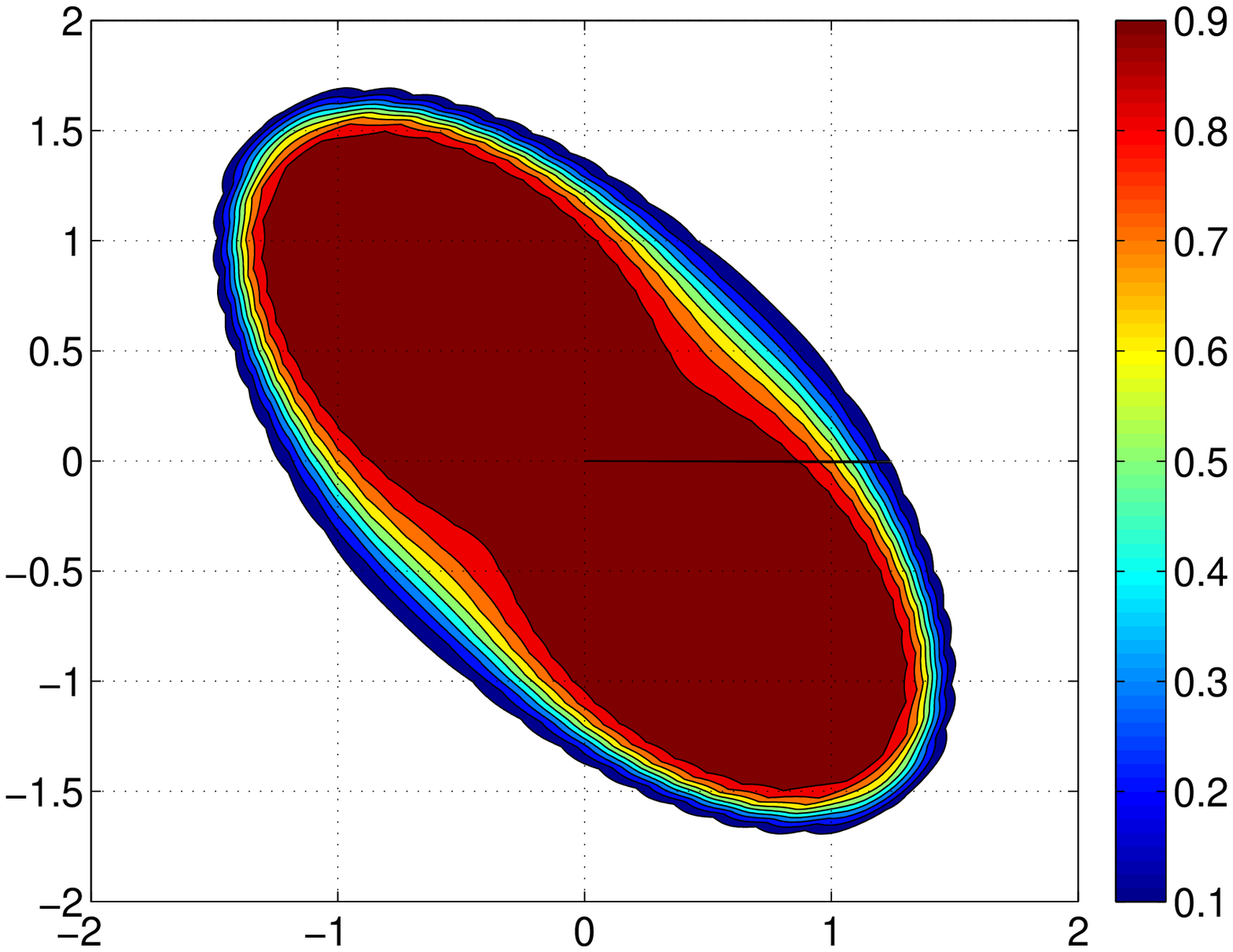}
\includegraphics[width=2in, height=2in]{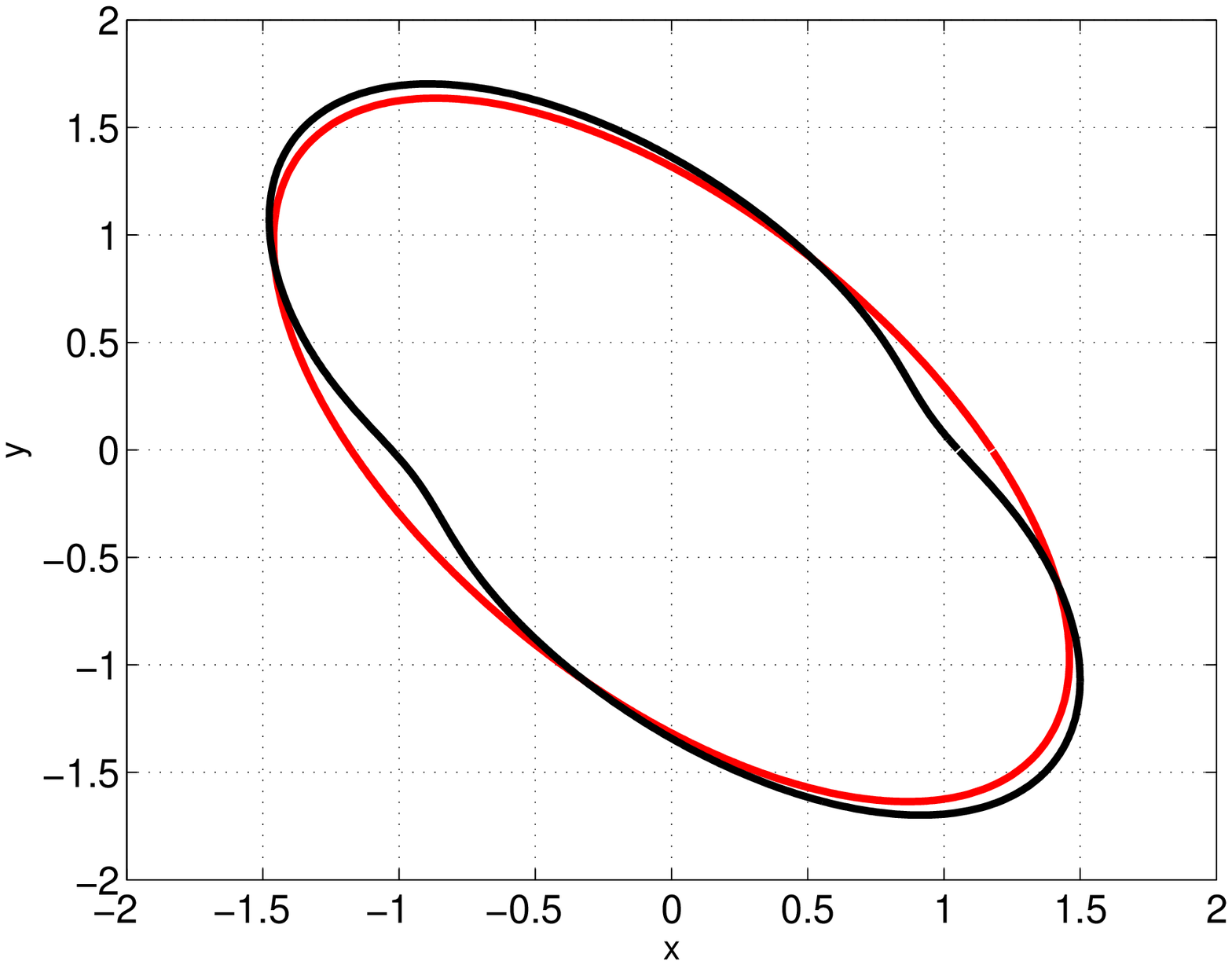}\\
\par
\smallskip
\end{center}
\par
\noindent\textsf{Figure 2. Solution of (24) obtained by perturbation for $\epsilon =0.05,\ \ M=1$. (a) Initial vortex ,  (b) Vortex at terminal moment ($T=1$)}, (c) Comparison vorticity line $w=0.3$ for Kirchhoff patch ($\epsilon=0$)  and the perturbed Kirchhoff patch corresponding to initial condition (25) with $\epsilon =0.05$.
\end{figure}

Assume that the equation $r=R(\varphi)$ represents an ellipse, i.e the solution of (24) with the initial elliptic patch $\Omega_0$  is the well known Kirchhoff vortex,  [12]. Any attempt to correct  that solution for small $\epsilon$ fails because the derivative of $\Omega_{\epsilon}(r,\varphi)$ at $\epsilon=0$ is infinite.  However, proceeding to
$$
\rho_{\epsilon}(\varphi,w)=\displaystyle R(\varphi)\left[S^{-1}\left(\frac{w}{M}\right)\right]^\epsilon , \eqno (26)
$$
obtained from (25) we arrive at an analytic function of $\epsilon$. That allows for application of a standard perturbation approach  with details given in Appendix. 

We show results in Fig. 2 from which one can see that the vorticity lines are loosing elliptic shape, nevertheless preserving the central symmetry. The latter obviously follows from the original equation (1).  Thus, the integral on the right hand side of (21) is zero and the vortex center does not move . 

The only purpose of the example was to illustrate a well-posedness of the suggested perturbation procedure.

Now we consider a dipole determined by $N=2,\ \ \ \omega_1=M>0,\ \omega_2=m<0 $.  From (17) and the general translation equation (next to (19)) one gets

\bigskip

{PROPOSITION 6. } 
$$ 
\begin {array}{ccc}
\displaystyle\frac 14\frac{\partial\rsf}{\partial t}=  
\frac{\partial}{\partial\varfi}\left\{\frac{\delta H}{\delta\rsf}-\rho_1(\varphi,w)\left(D_\varphi\frac{\delta H}{\delta\rho_1^2(\varphi,w)}\right)_{w=M}\right\}\ ,\\ \\  
\displaystyle\frac 14\frac{\partial\rss}{\partial t}= 
\frac{\partial}{\partial\varfi}\left\{\frac{\delta H}{\delta\rss}-\rho_2(\varphi,w)\left(D_\varphi\frac{\delta H}{\delta\rho_2^2(\varphi,w)}\right)_{w=m}\right)\ ,
\end{array}
\eqno (27)
$$  
$$  
\begin{array}{l} 
\displaystyle\frac{\partial z_1^*}{\partial
t}=\displaystyle\frac{1}{2\pi i}\int_0^{M}\int_0^{2\pi}\rho_1(\varphi, w)e^{-i\varphi} d\varphi
dw+ \frac{1}{2\pi i}\displaystyle\int_{m}^0\int_0^{2\pi}  
\frac{w\rho_2(\varphi, w)\left(\partial\rho_2(\varphi, w)/\partial w\right)dwd\varphi }{z_1-z_2-\rho_2(\varphi,
w)e^{i\varphi}}\ ,\\ \\ 
\displaystyle\frac{\partial z_2^*}{\partial
t}=\displaystyle\frac{1}{2\pi i}\int_{m}^0\int_0^{2\pi}\rho_2(\varphi, w)e^{-i\varphi} d\varphi
dw  
+\frac{1}{2\pi i}\displaystyle\int_0^{M}\int_0^{2\pi}
\frac{w\rho_1(\varphi, w)\left(\partial\rho_1(\varphi, w)/\partial w\right)dwd\varphi}{z_2-z_1-\rho_1(\varphi,
w)e^{i\varphi}}\ , 
\end{array} 
$$  
{\it {where an expression for Hamiltonian in terms of }} $\rho_1, \rho_2, \xi_1, \eta_1, \xi_2, \eta_2$ {\it {can be obtained from Proposition 5}} 

\smallskip

Evolution and translation equations from Proposition 6 were first announced  in [13]. Here they are essentially simplified and corrected.

Notice that for vortices of identical shape $\rho_1\equiv \rho_2$ the distance between poles and the angle between the axis through poles and the $x$-axis are conserved $\partial({\bf {z}}_1+{\bf {z}}_2)/\partial t=0$, the fact well known in contour dynamics.
\begin{figure}[tbph]
\begin{center}
\includegraphics[width=1.3in, height=1.3in]{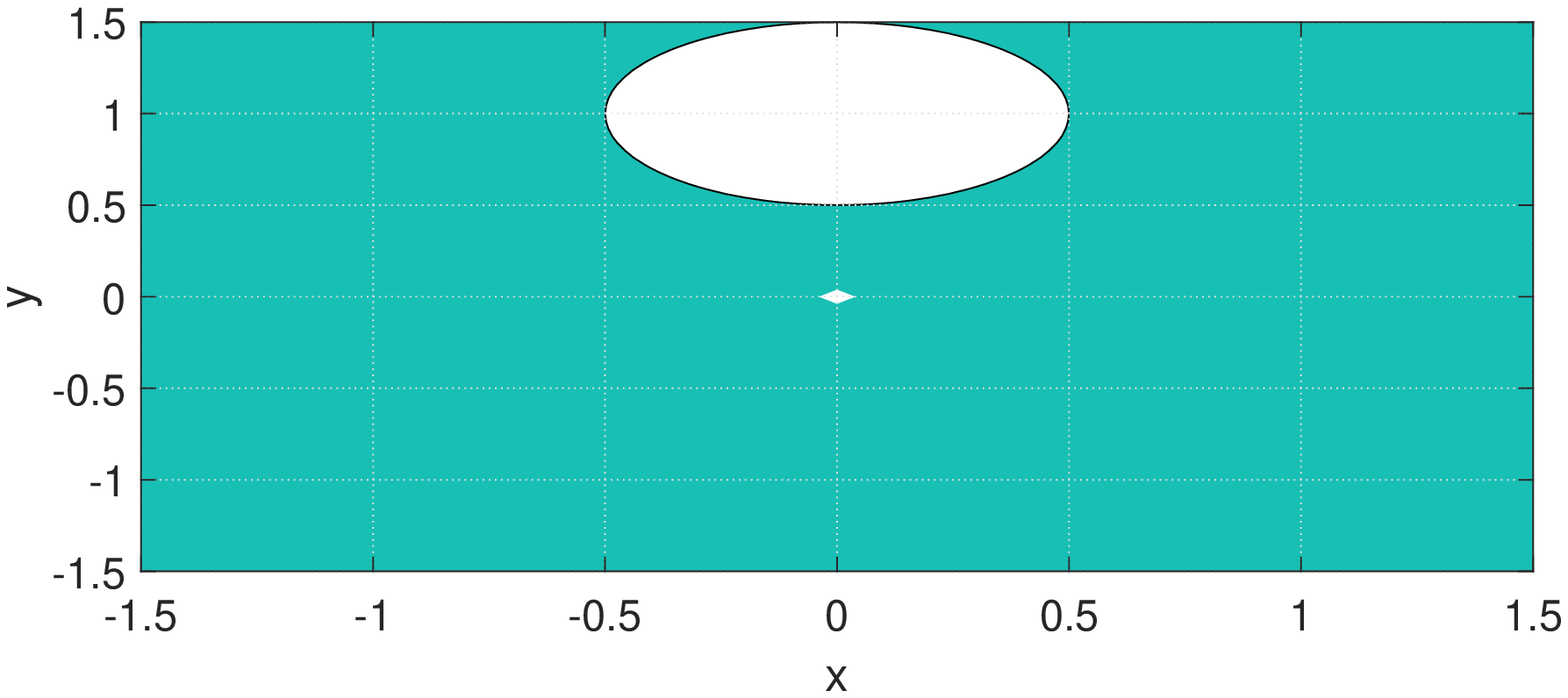} %
\includegraphics[width=1.3in, height=1.3in]{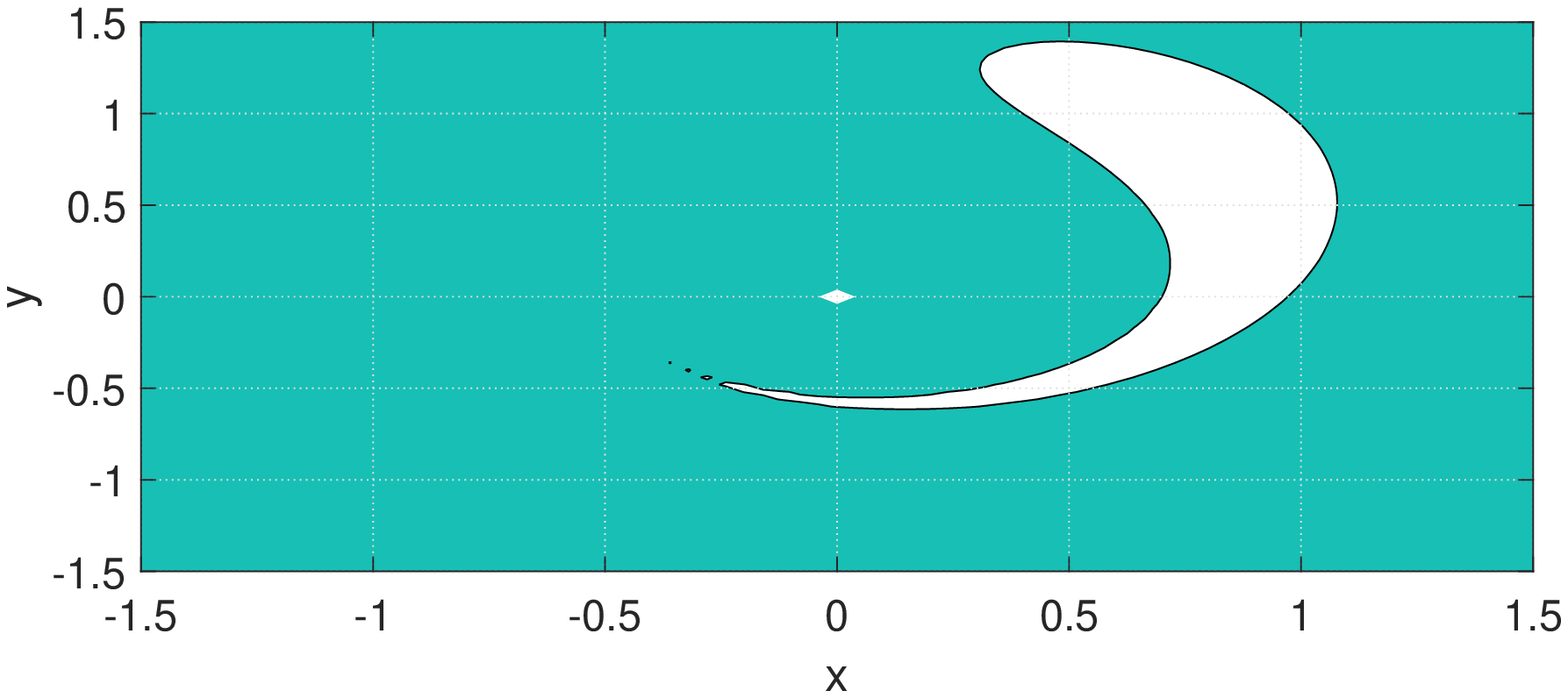}
\includegraphics[width=1.3in, height=1.3in]{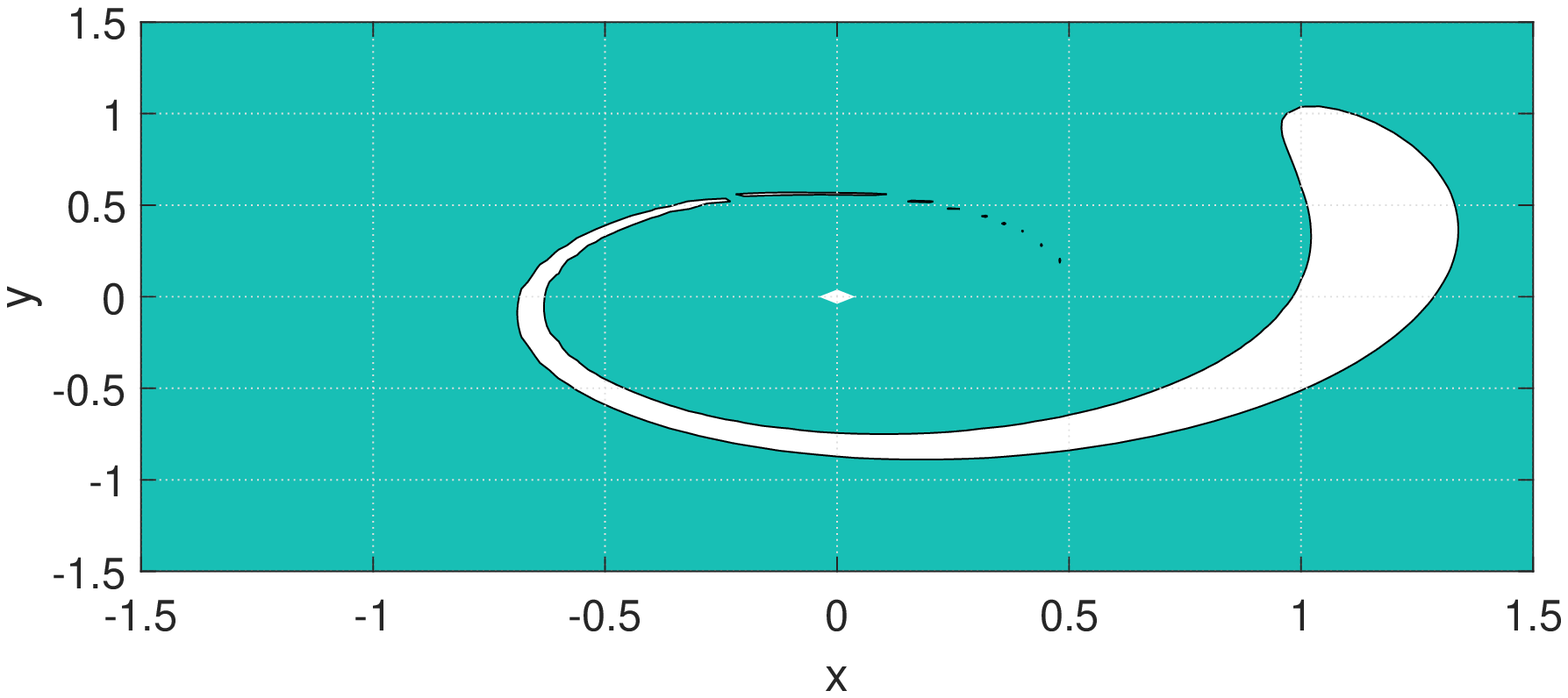}
\includegraphics[width=1.3in, height=1.3in]{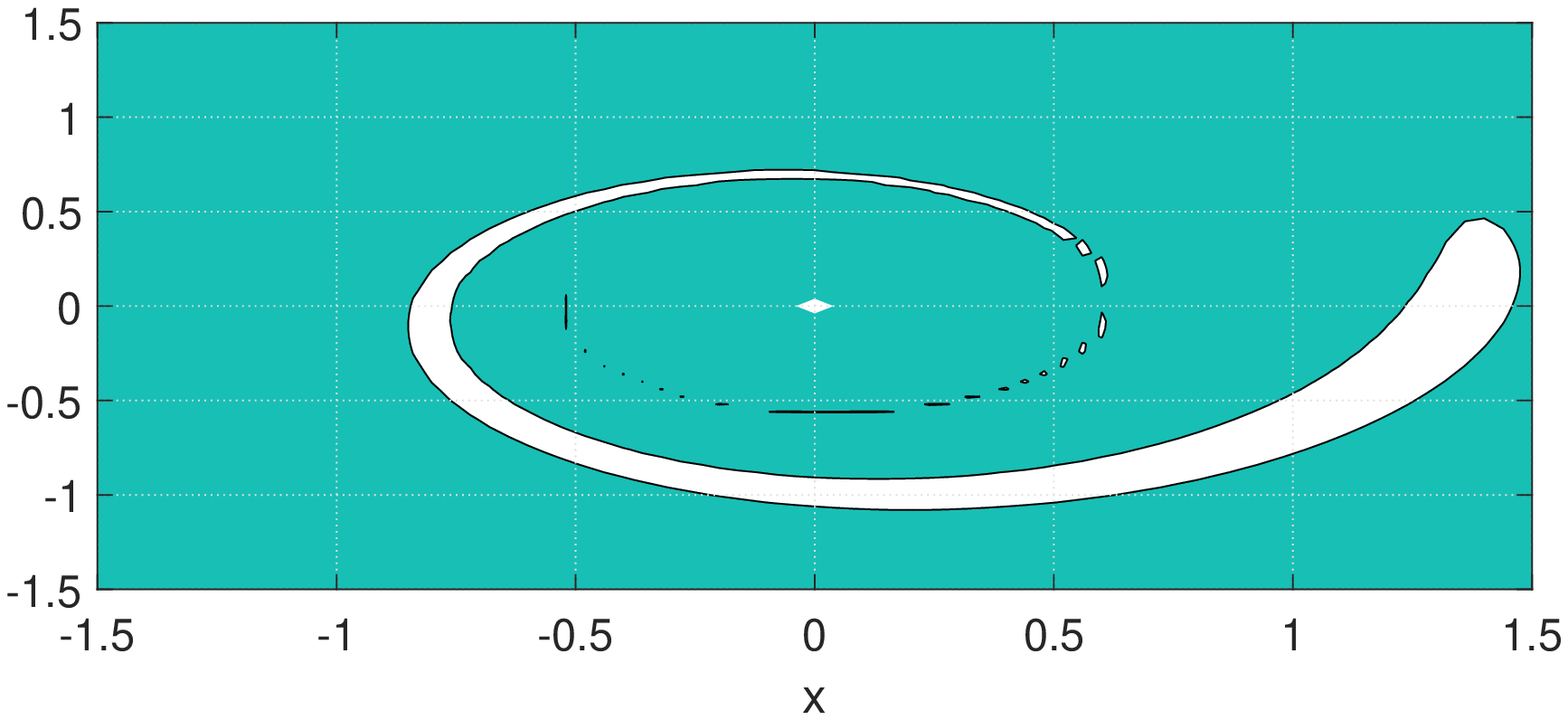}
\includegraphics[width=1.3in, height=1.3in]{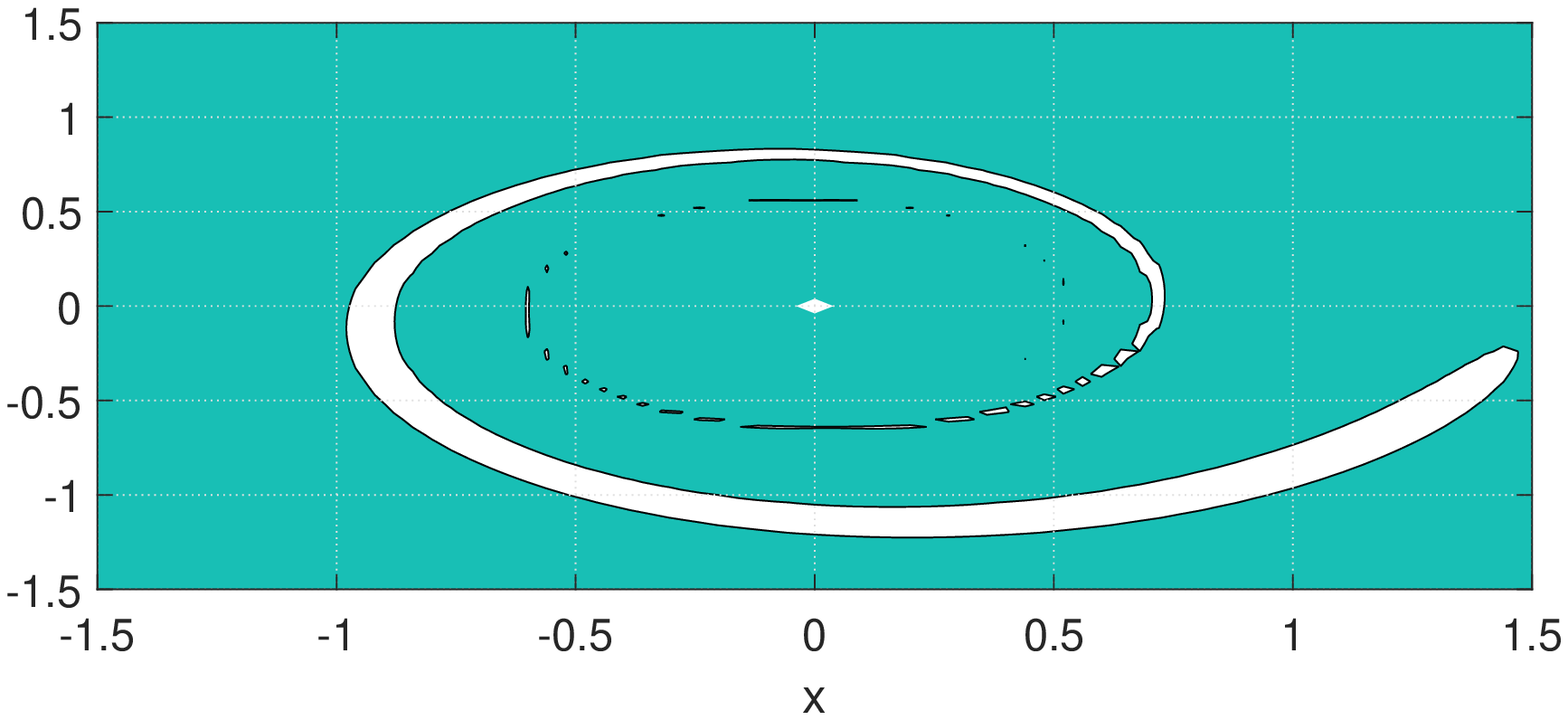}\\
\end{center}
\smallskip
\par
\noindent\textsf{Figure 3. Evolution of a weak satellite in a velocity field generated by a stationary point vortex at different time moments }
\end{figure}

To illustrate the use of (27) let us consider  a strong positive point vortex with maximum $M$ centered at the origin and a weak negative satellite  with the minimum $m,\ \ |m |\ll M, $ initially  axisymmetric and centered at $(0,R)$.  We assume that the stream function $\psi_1(r)$ of the positive vortex is not affected by the satellite. In addition we neglect the influence of the velocity field generated by the the satellite on itself. In other words, it is considered as a passive scalar driven by the velocity field of the strong vortex.  Hence, the first equation in (28) turns to $\rho_1(t,\varphi, w)= \rho_1(0,\varphi, w)$  and the second one turns to a closed equation 
$$  
\frac 12\frac{\partial\rss}{\partial t}= 
-\frac{\partial}{\partial\varfi}\psi_1\left(\rho_2(\varphi, w)^2+R^2-2R\rho_2(\varphi, w)\cos\varphi\right) .
$$ 
Notice that $w$ is included in the equation simply as a parameter.
Proceeding to the polar coordinates $(r,\theta)$ with a pole at the origin $(0,0)$, i.e. setting $r^2= \rho_2(\varphi, w)^2+R^2-2R\rho_2(\varphi, w)\cos\varphi,\ \ \sin\theta=\rho_2\sin\varphi/r$ obtain
$$
\displaystyle\frac{\partial r}{\partial t}+\frac 1r\frac{\partial \psi_1(r)}{\partial r}\frac{\partial r}{\partial\theta} =0 .
$$
The equation is integrable for any $\psi_1(r)$, but it makes a physical sense only if the background vortex is a point vortex. i.e. $\psi_1=k\ln r$ , where $k$ is its intensity. The solution is given by
$$
\displaystyle r^2-2rR\cos\left(\theta-\frac{kt}{r^2}\right)= C(w) , \eqno (28)
$$
where constant $C(w)$ is defined by the vorticity level $w$ and shape of the initial satellite.

It can be shown that the vorticity line of level $w$ spirals into limit cycle $r=R-r_0(w)$ where $r_0(w)$ is the radius  of the $w$-vorticity line for the initial satellite and the number of cycles in spiral $ n\approx c(w)\omega_0 t$, $\omega_0=k/R^2$ angular velocity, $t\ $ time,  $c$ constant depending on vorticity level.

 Notice that the integral (28) follows directly from the original equation (1) after linearization, [14]. The mentioned work addressed physical aspects of the solution and validation of the considered approximation. Moreover, a general case of a distributed background vortex was also discussed in [14].

\bigskip

{\large {\bf{5.   Discussion and  Conclusions}}}

\smallskip

We have introduced a class of vorticities extending  the contour dynamics[4,5] to the case of continuously distributed vorticity and developed a Hamiltonian formalism for that class.
The relation with contour dynamics was revealed and discussed in the text.  In that regard the suggested approach could be called "continuum contour dynamics".

 In this section we, first,  present explicit scalings transforming the suggested class to two  well known models, the point vortex system, [11], and FAVOR, [6].

As for the former, let $\Omega_k(\br)=\Omega(\br)I_{G_k}(\br)=\tilde \Omega_k(\br-{\bf{z}}_k)$ be the $k$-th vortex written in the local coordinate system with origin at ${\bf{z}}_k$, then
$$
\Omega_{\epsilon}(\br)=\displaystyle \frac{1}{\epsilon^2}\sum_k\tilde \Omega_k\left(\frac{\br-{\bf{z}}_k}{\epsilon}\right)\longrightarrow \sum_k\bar {\omega}_k\delta(\br-{\bf{z}}_k) ,
$$
as $\epsilon \to\ 0$, where
$$
\bar {\omega}_k=\displaystyle \frac 12\int_0^{2\pi}\int_0^{\omega_j}\rho_k^2(\varphi,w)dwd\varphi.
$$
For the latter,  let
$$
\displaystyle R_k^2(\varphi)=-\left. |\omega_k|\frac{\partial\rho^2}{\partial w}\right|_{w=\omega_k}
$$
be a a characteristic space scale near the $k$-th vortex peak. Obviously $r=R_k(\varphi)$ is an ellipse and  vorticity lines $\Omega_k=w$ with $w$ close to $\omega_k$ are well approximated by  $r=c(w)R_k(\varphi)$ with a constant depending on the vorticity level. Introduce dimensionless distance $\tilde r=r/R_k(\varphi)$ and define 
$\tilde \Omega_k(\varphi, \tilde r)=\Omega_k(\varphi, r)$ where $(\varphi, r)$ are local polar coordinates. Introduce
$$
\Omega_{\epsilon}(\br)=\displaystyle \sum_k\tilde\Omega_k\left(\varphi, \tilde r^{1/\epsilon}   \right) ,
$$
 then
$$
\Omega_{\epsilon}(\br)\displaystyle \longrightarrow \sum_k\bar {\omega}_kI_{E_k}(\br)
$$
as $\epsilon \to\ 0$, where
$$
E_k=\displaystyle \left\{ (\varphi, r)|\ \ r<R_k(\varphi)  \right\}
$$ is a  Kirchhoff elliptic patch, [12].

The described limiting procedures lead to the well known Hamiltonian formulations of the point vertex system, [2], and FAVOR, [15]. However, details of a transition from Proposition 5  to equations presented in in the mentioned papers are out of the scope of our present work.

Then, notice that most of the above results can be extended to a similar  class of vortices on an arbitrary two-dimensional Riemann manifold  $\texttt{M}$  with metric $ dS=s(x,y)dxdy, (x,y)\in D,
$  where $D$ is a region in the $(x,y)$-plane, $s=s(x,y)$ is the  metric density.  This is possible because  the vorticity conservation equation on $\texttt{M}$ similar to (1)
$$
\frac{\partial\Omega}{\partial t}+s^{-1}J(\psi, \Omega)=0 
$$  
can be written in the Hamiltonian form  as well
$$
\frac{\partial q}{\partial t}=\{q,H\}\ ,
$$  
where $q=s\Omega$ is the phase variable  and the non-canonical Poisson bracket is expressible similarly to (2) as
$$  
\{F,G\}=\int_D\Omega(\br)J_{x,y}\left(\frac{\delta F}{\delta q(\br)}, \frac{\delta
G}{\delta q(\br)}\right)d\br\ ,  
$$  
with Hamiltonian $  H=-\frac 12\int_Dq\psi d\br $. The ideal hydrodynamics in the plane is given by $s\equiv 1,\ \ D=R^2$. For a sphere of unit radius $x=\lambda$ is the longitude, $y=\theta$ is the latitude and $D=[0,2\pi]\times[0,\pi]$, and, finally, for periodic boundary conditions on a rectangle $[0,a]\times[0,b]$ $\texttt{M}$ in a torus, $s\equiv 1/(ab),\ \ D=[0,2\pi]\times[0,2\pi]$
Dipoles on a sphere were discussed in [13].

Finally, summing up the results presented we conclude  that, from the application point of view,  there is no convincing evidence yet that the equations in terms of vorticity lines and coordinates of extrema could be more efficient in analytical/numerical studies of vortex systems than  traditional approaches. However, certain similarities with the contour dynamics, a theory proved to be useful, give hopes  for a better future.

Theoretically, the suggested approach gives a useful insight in conservation laws concerning with the vorticity topography. Yet, a traditionally interesting problem of a Poisson bracket diagonalization is solved for the suggested class of vorticities. However, we admit that the diagonalization here was a goal itself unlike a similar procedure for Hasegawa-Mima equation , [15,16] that has led to canonical variables and ultimately to advances in the weak turbulence theory. Technically, canonical variables could be introduced for the Hamiltonian system discussed here as well, but they would hardly make a clear physical sense.

\bigskip

  {\large {\bf{6.  Appendix}}}
	
	\bigskip

	{\bf{6.1\ Proof of Lemma }} 
	
	\smallskip

Identity $p(\hat{x}(p,w),\hat{y}(p,w))=p$	after differentiation in $p$ and $w$ gives $p_x=\hat{y}_w/g,\ p_y=-\hat{x}_w/g $. The same identity implies
$$
\displaystyle p_x\frac{\delta\hat{x}}{\delta\Omega(\br)}+p_y\frac{\delta\hat{y}}{\delta\Omega(\br)}=0\ \ {\rm {or}}\ \  \hat{y}_w\frac{\delta\hat{x}}{\delta\Omega(\br)}-\hat{x}_w\frac{\delta\hat{y}}{\delta\Omega(\br)}=0\ . \ \eqno (A1)
$$
Taking variational derivatives in identity $\Omega (\xi+(\hat{x}(p,w), \eta+(\hat{y}(p,w))=w$ obtain
$$
\displaystyle \delta\Omega(\br)+\Omega_x(\delta\xi+\delta\hat{x})+\Omega_y(\delta\eta+\delta\hat{y})=0\ .\ \eqno (A2)
$$
From $\Omega(x,y)=w$ it follows that $\Omega_x=-\hat{y}_p/g,\ \ \Omega_x=\hat{x}_p/g$. Plug these expressions in (A2) and solve (A1-A2) for $\delta \hat{x}/\delta\Omega(\br)$ and $\delta \hat{y}/\delta\Omega(\br)$. The result is
$$
\begin{array}{cc}
\displaystyle \frac{\delta \hat{x}(p,w)}{\delta\Omega(\br)}= \frac{\hat{x}_w(p,w)}{g(p,w)}\left(\delta(\Omega(x,y)-w)\delta(p(x,y)-p)-\hat{y}_p\frac{\delta \xi}{\delta\Omega(\br)}-\hat{x}_p\frac{\delta \eta}{\delta\Omega(\br)}\right),\ \ \frac{\delta \hat{y}}{\delta\Omega(\br)}=\frac{\delta \hat{x}}{\delta\Omega(\br)}\frac{\hat{y}_w}{\hat{x}_w} .
\end{array}
$$
Substituting expressions for the variational derivatives of $\xi$ and $\eta$ on page 4, converting them to the delta functions in $x$ and $y$, arrive at (12).
\bigskip

{\bf{ 6.2  Perturbation procedure}}

\smallskip 

Assume for the initial condition in (24)
$$
p(\varphi,w)=p(\varphi,w;\epsilon)=p_0(\varphi,w) +\epsilon p_1(\varphi,w)+...
$$
and represent the solution  in a similar way
$$
\rho^2(\varphi,w)=\rho^2_0(\varphi,w) +\epsilon\rho^2_1(\varphi,w)+...
$$
and get
$$
\displaystyle \frac{\partial}{\partial t}\rho^2_n=L_{n-1}(\rho^2_n),\ \ \rho^2_n|_{t=0}=p_n(\varphi,w) , \eqno(A3)
$$
where
$$
L_{n-1}(\rho^2)=\displaystyle \left. \frac{\delta N (\rho^2)}{\delta\rho^2}\right|_{\rho=\rho_{n-1}}
$$
is the linearization of $N$ at the previous correction.

Take the initial condition in form (26) and denote 
$$
f(w)=\ln S^{-1}\left(\frac{w}{M}\right) .
$$
Thus, in the first order of $\epsilon$
$$
p(\varphi,w)=R^2(\varphi)+2\epsilon R^2(\varphi)f(w)
$$
Because $p_0$ does not depend on $w$,  the zero approximation $\rho^2_0=\rho_0^2(t,\varphi)$ does not depend on $w$ as well and in fact  is nothing but the solution of the contour dynamics equation corresponding to the initial condition at $\epsilon =0$. In addition, assume that $R(\varphi)$ is symmetric about the pole thereby the singular term in (24) vanishes ($\psi_\xi=\psi_\eta=0$), i.e. the vortex center does not move. That implies  essential simplifications in the linearized equation (A3) for $n=1$
$$
\displaystyle \frac{\partial}{\partial t}\rho^2_1(\varphi, w)= -\int_0^{2\pi}K(\varphi,\theta)\int_0^{M}\rho_1^2(\theta,u)dud\theta,\ \ \left. \rho^2_1\right |_{t=0}=2R^2(\varphi)f(w) , \eqno(A4)
$$
where the kernel is
$$
\displaystyle K(\varphi,\theta)=\frac{1}{2\pi }\frac{\partial}{\partial\varphi}\ln \left(\rho_0^2(\varphi)+\rho_0^2(\theta)-2\rho_0(\theta)\rho_0(\varphi)\cos(\theta-\varphi)\right) .
$$
Integrate both parts of (A4) in $w$ and get
$$
\displaystyle \frac{\partial}{\partial t}z(\varphi)= -M\int_0^{2\pi}K(\varphi,\theta)z(\theta)d\theta,\ \ \ z|_{t=0} = 2\bar fR^2(\varphi) ,
$$
where
$$
z(\varphi)=\displaystyle \frac{1}{M}\int_0^M \rho^2_1(\varphi, w)dw,\ \ \bar f= \frac{1}{M}\int_0^M f(w)dw .
$$
This equation is easy to solve numerically and then retrieve $\rho_1(t,\varphi, w)$ itself using the initial  condition given in (A4). The result is
$$
\rho^2_1(t,\varphi, w)=z(t,\varphi)+2R^2(\varphi)(t,\varphi)\left(f(w)-\bar f\right) .
$$
Converting
$$
\rho^2(t,\varphi, w)=\rho^2_0(t,\varphi)+\epsilon \rho^2_1(t,\varphi, w)
$$
with respect to $w$, we obtain the first order approximation for the vorticity itself
$$
\Omega_{\epsilon}(\xi+r\cos\varphi, \eta+r\sin\varphi)=\displaystyle MS\left(\exp\left\{\frac{r^2-\rho_0^2(t,\varphi)-\epsilon(z(t,\varphi)-z(0,\varphi))}{2\epsilon R^2(\varphi)}\right\}\right) .
$$

\bigskip

{\large \textbf{Acknowledgment}}

\smallskip

The author is very grateful to George Sutyrin for a useful discussion on the dipole example. Comments of an anonymous reviewer helped to correct numerical results concerning with monopole.

%
%
%
%

\pagebreak

{\large \textbf{References}}

\smallskip

1. V.E.Zakharov and E.A. Kuznetsov, {\it{Soviet Sci. Rev. Sect. C: Math. Phys. Rev.}}, 4 (1984), 167-220

2. V.P. Goncharov, V.I. Pavlov, The problems of hydrodynamics in Hamiltonian description, {\it Moscow State University}, (1993) 

3. V.I. Arnold, Mathematical Methods of Classical Mechanics, {\it{Springer, Berlin}} (1978).
		
4. G.S. Deem, N.J. Zabusky,   {\it{ Phys. Rev. Lett.}} {\bf{40}} (1978) 859-877.
  
5. V.P. Goncharov, V.I. Pavlov, {\it{Mathematical Reviews}}, (2001)179-237

6. M. V. Melander, A. S. Styczek, and N. J. Zabusky,  {\it{Phys. Rev. Lett.}}, {\bf{53}},   (1984), 1222-1245

7. R.J. Trudeau, Introduction to Graph Theory, {\it{Kent State University Press}}, (1976)

8.  P. Meunier, S. Le Dizès, T. Leweke,   {\it{C. R. Physique }} {\bf{6}} (2005), 431–450

9. A.S. Monin, A.M. Yaglom, Statistical Fluid Mechanics, Volume 2,  Mechanics of Turbulence, {\it{ MIT Press}}, (1975)

10.  L.I. Piterbarg ,  {\it{Physics Letters A}}, {\bf{205}} (1995), 149-157

11. H. Aref,   {\it{Annual Review of Fluid Mechanics}}, {\bf{ 15}} (1983), 345-389.

12.  G.R.Kirchhoff, Vorlsungen uber Mathematische Physik, Vol.1, {\it{Leipzig, Teubner}}, (1876) 466 pp. 

13. L.I. Piterbarg,  {\it Physics Letters A}, {\bf 372} (2008) 2032-2038

14. G. Sutyrin,  {\it Phys. Fluids} {\bf 31},  (2019), 103-111

15.  M.A. Brutyan and P.L. Krapivskii,  {\it PMM U.S.S.R}, {\bf 52}:1, (1988), 130-132

16.  V.E.Zakharov and L.I.Piterbarg,  {\it Phys. Lett., A}, {\bf 126}:8, (1988), 497-501

17.  L.I. Piterbarg, {\em Amer. Math. Society Transl.}, {\bf 182} (1998) 131-65.

\end{document}